\documentclass{article}
\usepackage[dvipdfmx]{graphicx}
\usepackage{
spconf,
amsmath,
amsfonts,
cite,
xurl,
color,
mathrsfs,
multirow,
array,
colortbl
}

\usepackage{algorithmic}
\usepackage{algorithm}

\usepackage{mathrsfs}

\newlength{\tablelength}
\newcolumntype{C}[1]{>{\centering\arraybackslash}p{#1\tablelength}}
\newcolumntype{L}[1]{>{\arraybackslash}p{#1\tablelength}}

\newcommand{\bm}[1]{\boldsymbol{#1}}

\newcommand{\mysection}[1]{
\vspace{-5pt}
\section{#1}
\vspace{-3pt}
}
\newcommand{\mysubsection}[1]{
\vspace{-5pt}
\subsection{#1}
\vspace{-3pt}
}

\newcommand{\mysubsubsection}[1]{
\vspace{-4pt}
\subsubsection{#1}
\vspace{-3pt}
}

\title{
Real-time speech enhancement using equilibriated {RNN}
}

\name{Daiki Takeuchi$^\dagger$, Kohei Yatabe$^\dagger$, Yuma Koizumi$^\ddag$, Yasuhiro Oikawa$^\dagger$, Noboru Harada$^\ddag$
\vspace{-3pt}}
\address{
$^\dagger${\fontsize{11pt}{0pt}\selectfont Department of Intermedia Art and Science, Waseda University, Tokyo, Japan}\\
$^\ddag${\fontsize{11pt}{0pt}\selectfont NTT Media Intelligence Laboratories, Tokyo, Japan}
}

\begin{document}
\ninept
\maketitle

\begin{abstract}
We propose a speech enhancement method using a causal deep neural network~(DNN) for real-time applications.
DNN has been widely used for estimating a time-frequency~(T-F) mask which enhances a speech signal.
One popular DNN structure for that is a recurrent neural network~(RNN) owing to its capability of effectively modelling time-sequential data like speech.
In particular, the long short-term memory (LSTM) is often used to alleviate the vanishing/exploding gradient problem which makes the training of an RNN difficult.
However, the number of parameters of LSTM is increased as the price of mitigating the difficulty of training, which requires more computational resources.
For real-time speech enhancement, it is preferable to use a smaller network without losing the performance.
In this paper, we propose to use the equilibriated recurrent neural network~(ERNN) for avoiding the vanishing/exploding gradient problem without increasing the number of parameters.
The proposed structure is causal, which requires only the information from the past, in order to apply it in real-time.
Compared to the uni- and bi-directional LSTM networks, the proposed method achieved the similar performance with much fewer parameters.
\end{abstract}
\begin{keywords}
Real-time speech enhancement, equiribriated recurrent neural network, vanishing/exploding gradient problem.\vspace{2pt}
\end{keywords}

\mysection{Introduction}
Speech enhancement is used for recovering the target speech from a noisy observed signal. 
In the single-channel case, the standard method is time-frequency (T-F) masking in the short-time Fourier transform (STFT) domain.
Recently, speech enhancement is advanced by the use of a deep neural network (DNN) to estimate a T-F mask.
For effectively modelling a speech signal which is time-sequential data, a recurrent neural network~(RNN) is used in various speech signal processing applications \cite{
Wang2018supervised
,Erdogan2015phase
,hershey2016deep
,Zhao2018convolutional
,kalchbrenner2018efficient
,perotin2918multi
,LeRoux2019Phasebook
,takeuchi2019data
,koizumi2019trainable
,fu2019metricgan
,chakrabarty2019time
,zheng2019phase
,Koizumi2020Speech
,Kawanaka2020stable
}.

\begin{figure}[t]
  \centering
  \vspace{-2pt}
   \includegraphics[width=0.96\columnwidth]{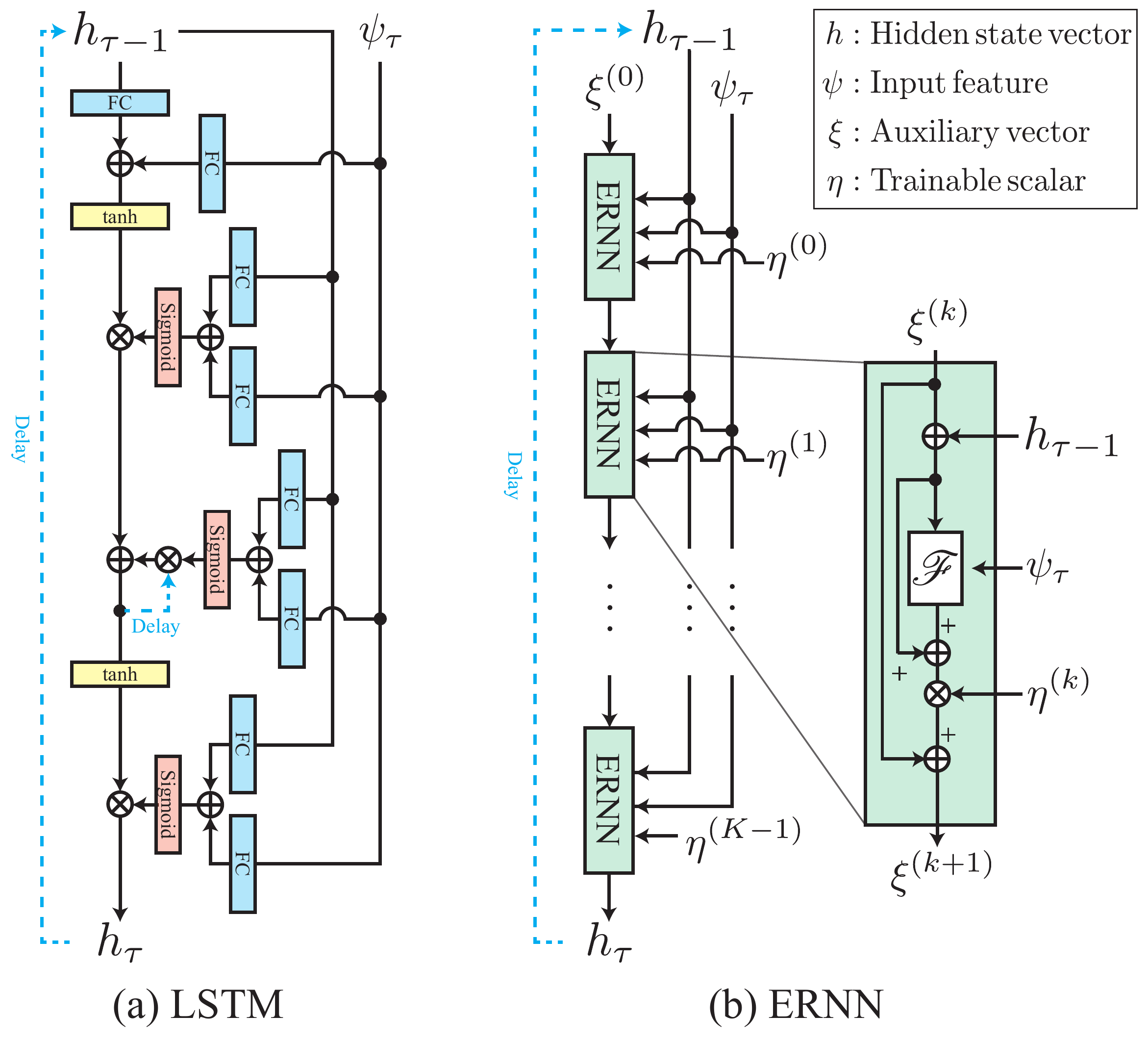}
  \vspace{-10pt}
  \caption{Block diagrams of LSTM and ERNN. ``FC" stands for fully-connected layer. 
  The same DNN $\mathscr{F}$ is repeatedly applied in ERNN.}
  \label{fig:LSTM_ERNN}
  \vspace{-2pt}
\end{figure}

While it has been effectively applied to speech enhancement, RNN is difficult to train in general because the gradient of RNN vanishes or explodes at an exponential rate by performing back-propagation to the same layer repeatedly.
This difficulty of training RNN is so-called the vanishing/exploding gradient problem \cite{bengio1994learning}, and several methods have been proposed to solve it \cite{Arjovsky2016unitaryRNN,Wisdom2016uniRNNforSpeech,Hochreiter1997LSTM}.
One of the popular DNN structures to mitigate this problem is the long short-term memory~(LSTM) \cite{Hochreiter1997LSTM} illustrated in Fig.~\ref{fig:LSTM_ERNN}(a).
By combining three gated units (input gate, forget gate and output gate), LSTM solves the vanishing gradient problem to some extent.
As it can be trained effectively in practice, LSTM and the bidirectional LSTM~(BLSTM) has been applied to speech enhancement and performed better than the conventional methods at the time \cite{
Erdogan2015phase
,Zhao2018convolutional
,takeuchi2019data
,koizumi2019trainable
,fu2019metricgan
,chakrabarty2019time
,zheng2019phase
,Koizumi2020Speech
,Kawanaka2020stable
}.

Considering a practical situation in the real world, some research on DNN-based speech enhancement has focused on real-time application \cite{
Naithani2017lowLatency
,Tan2018convRNN
,parviainen2018time
,Pandey2019TCNN
,Bhat2019realTime}.
To apply an enhancement method in real time, the system must be causal, i.e., it uses past information only and does not require future information to estimate the enhanced signal.
Therefore, uni-directional LSTM are often used in that task \cite{Naithani2017lowLatency,Tan2018convRNN,parviainen2018time,Pandey2019TCNN}.
However, as the price of mitigating the vanishing gradient problem, LSTM consists of a lot of parameters as in Fig.~\ref{fig:LSTM_ERNN}(a).
Since more parameters require more computational resources, a simpler RNN should be more suitable for real-time speech enhancement than LSTM if the 
gradient problem can be solved in a different way.

In this paper, we propose a real-time speech enhancement method using a causal RNN with much fewer parameters compared to LSTM.
In the proposed method, the equilibriated recurrent neural network (ERNN) \cite{kag2019rnns} is used for the T-F mask estimator.
ERNN is a simpler RNN as in Fig.~\ref{fig:LSTM_ERNN}(b) and can avoid the vanishing/exploding gradient problem by iteratively applying the same layer to the hidden state vector.
Ideally, the gradient of ERNN in back-propagation does not vanish or explode \cite{kag2019rnns}, and therefore long-term dependencies in the sequential data can be learned without the gated units as in LSTM.
As a result, the number of parameters of ERNN can be noticeably decreased while maintaining the speech enhancement performance.
The experimental results confirmed that the proposed method can reduce the number of parameters to less than $1/5$ times that of the LSTM network without sacrificing the performance.

\vspace{-3pt}
\mysection{DNN-based speech enhancement}

This paper focuses on T-F masking for speech enhancement.
In this section, after introducing DNN-based T-F masking and RNN briefly, real-time speech enhancement is explained.

\mysubsection{Time-frequency masking based on DNN}
The aim of speech enhancement is to recover the target speech signal $s_t$ degraded by noise $n_t$ from an observed signal $x_t$,
\begin{equation}
    x_t = s_t + n_t,
    \label{eq:sse_td}
\end{equation}
where $t$ is the time index.
It can be rewritten in T-F domain as
\begin{equation}
    X_{\omega,\tau} = S_{\omega,\tau} + N_{\omega,\tau},
    \label{eq:sse_STFTd}
\end{equation}
where $X$ is the T-F representation of $x$ (spectrogram obtained by STFT in this paper), and $\omega = 1,\ldots,\Omega$ and $\tau = 1,\ldots,T$ denote the indices of frequency and time frame, respectively.
In T-F masking, the estimated target signal $\hat{S}_{\omega,\tau}$ is acquired by the element-wise multiplication of a T-F mask $G_{\omega,\tau}$ to the observation $X_{\omega,\tau}$: 
\begin{equation}
    \hat{S}_{\omega,\tau} = G_{\omega,\tau} \, X_{\omega,\tau}.
\end{equation}
Then, the enhanced result is transformed back to the time domain by the inverse transform.
The T-F mask $G_{\omega,\tau}$ must be estimated solely from $X_{\omega,\tau}$,
which is the difficult part.

Many methods have applied DNN to estimate the T-F mask.
In deep-learning-based approach, a T-F mask $G_{\omega,\tau}$ is estimated as 
\begin{equation}
     G_{\omega,\tau}=  \mathcal{M}_{\theta}(\Psi)_{\omega,\tau}
\end{equation}
where $\mathcal{M}_\theta$ is a regression function implemented by DNN, $\theta$ is the set of its parameters, and $\Psi = \Psi(X)$ is the input acoustic feature.
Since the signal is time-sequential data indexed by $\tau$, RNN is often used for realizing the regression function $\mathcal{M}_\theta$.

\mysubsection{Recurrent neural network (RNN) and LSTM}
Among many DNN structures, RNN is a popular network for modelling time-sequential data including speech.
RNN consists of a function $\mathscr{F}$ which output the current hidden state vector $h_\tau$ from the past state vector $h_{\tau-1}$ and the current input feature $\psi_\tau$ as follows:
\begin{equation}
    h_\tau = \mathscr{F}(\psi_\tau,h_{\tau-1}),
    \label{eq:simplestRNN}
\end{equation}
where the recurrent structure on the state vector $h_\tau$ enables to learn the long-term dependencies of time series with fewer parameters comparing to non-recurrent networks.

Although an RNN can effectively handle information from the past, it may not perform well in practice because of the difficulty on its training, the so-called vanishing/exploding gradient problem.
When back-propagation is performed to RNN, the gradient passes through the same layer repeatedly.
Then, by the chain rule, the gradient on the current state vector $h_c$ from the past state vector $h_p$ is the product of gradients for all intermediate state vectors:
\vspace{-4pt}
\begin{equation}
    \frac{\partial h_c }{\partial h_p} = \prod_{r = 0}^{c-p-1}
    \frac{\partial}{\partial h_{c-r-1}} \mathscr{F}(\psi_{c-r},h_{c-r-1}).
    \vspace{-4pt}
\end{equation}
Therefore, the back-propagated gradient vanishes or explodes at an exponential rate unless the norm of each gradient is equal to one, i.e., $\|\partial\mathscr{F}/\partial h_\tau\|=1\;\;(p\leq\tau<c)$.
Even though an RNN has ability to model the long-term dependency, learning it is difficult because the dependency between the current and past information is quickly lost as the gradient quickly vanishes.

To mitigate the vanishing or exploding gradient problem of RNN, several methods have been developed \cite{Arjovsky2016unitaryRNN,Wisdom2016uniRNNforSpeech,Hochreiter1997LSTM}.
One of the most standard methods is LSTM \cite{Hochreiter1997LSTM} illustrated in Fig.~\ref{fig:LSTM_ERNN}(a).
It includes an additional recurrent loop of the so-called cell state so that the information from the past is retained unless the forget gate eliminates it.
The magnitude of the gradient of LSTM does not decrease when the forget gate is open, and thus the vanishing gradient problem is avoided by assuming that the forgate gate properly select the information.
Since it works well in practice, LSTM plays an important role in the DNN-based speech enhancement.
However, since the gated unit consists of twice as many parameters than the linear layer, LSTM consists of a lot of parameters, which requires more computational resources compared to a simpler RNN.

\begin{figure}[t]
  \centering
  \includegraphics[width=0.98\columnwidth]{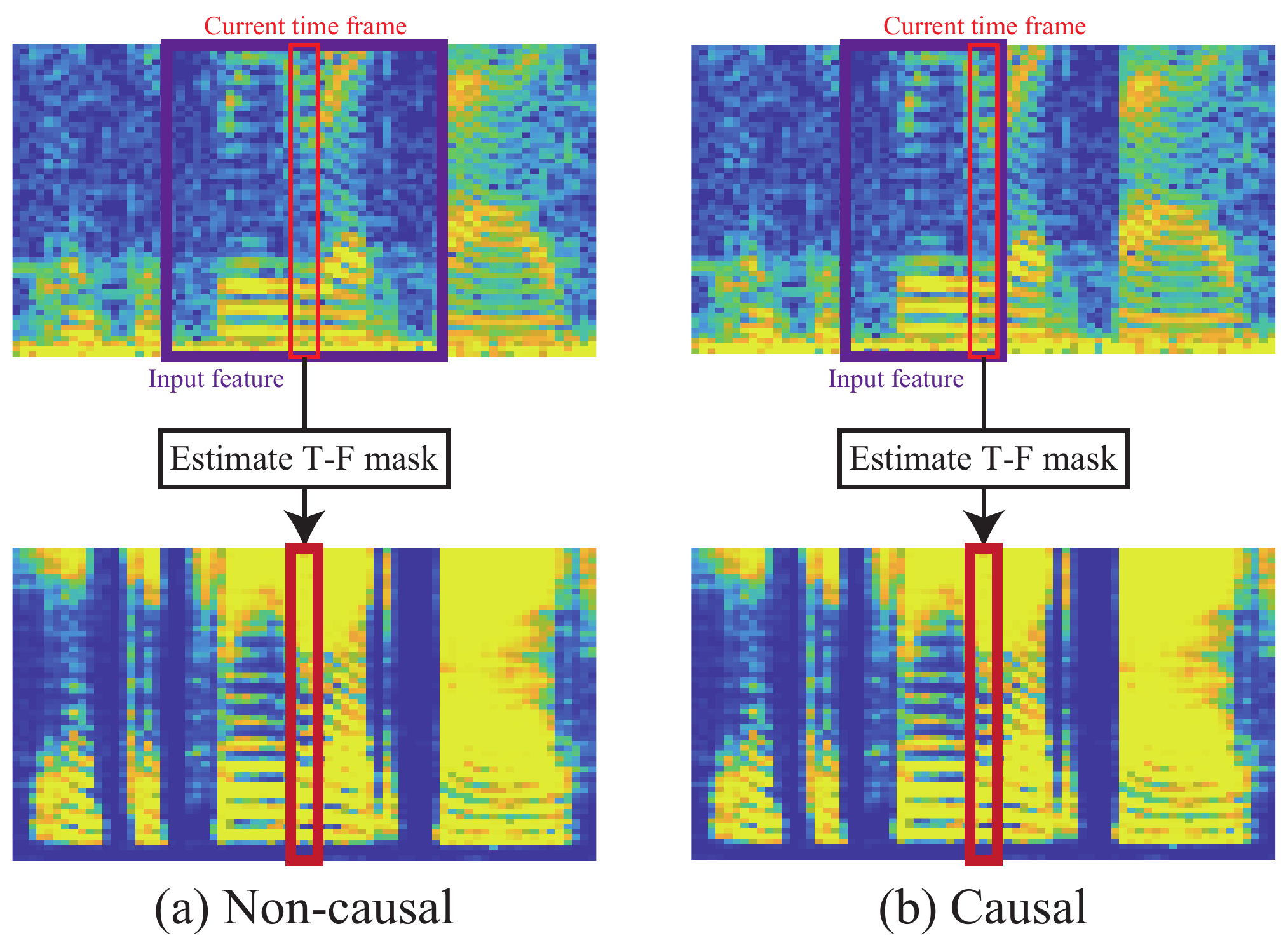}
  \vspace{-10pt}
  \caption{Illustration of non-causal and causal estimators. While a non-causal DNN uses future information for estimating the current T-F mask, a causal DNN requires only past and current information. }
  \label{fig:causality}
\end{figure}

\mysubsection{Real-time speech enhancement and causal RNN}
Some research on DNN-based speech enhancement has focused on the real-time application for applying it to a practical situation in the real world \cite{Naithani2017lowLatency,Tan2018convRNN,Pandey2019TCNN,Bhat2019realTime}.
To apply an enhancement method in real time, the system must be causal as illustrated in Fig.~\ref{fig:causality}.
In general, T-F mask $G$ at time index $\tau$ can be estimated from the input feature $\Psi$ obtained from both past and future,
\begin{equation}
    G_{\tau} = \mathcal{M}_\theta(\ldots, \psi_{\tau+1},\psi_{\tau},\psi_{\tau-1},\ldots),
\end{equation}
as illustrated in Fig.~\ref{fig:causality}(a), where $G_{\tau} = [G_{1,\tau},\ldots,G_{\Omega,\tau}]^\mathsf{T}$.
However, such non-causal network cannot be applied in real time because the information in future $\psi_{\tau+1},\psi_{\tau+2},\ldots$ is not available at the time of estimating $G_\tau$.
It requires some delay for buffering the input feature until all necessary information is obtained.
For real-time applications, the network must be causal, i.e., estimation must be performed based on the past information only:
\begin{equation}
    G_{\tau} = \mathcal{M}_\theta(\psi_{\tau},\psi_{\tau-1},\psi_{\tau-2},\ldots).
\end{equation}
This requirement makes RNN suitable for real-time applications because the past information can be encoded into the hidden state vector $h$ so that only the input feature $\psi_\tau$ and the state vector $h_{\tau-1}$ are required for estimating the mask at time $\tau$ as
\begin{equation}
    G_{\tau} = \mathcal{M}_\theta(\psi_{\tau},h_{\tau-1}).
    \label{eq:causalTFmask}
\end{equation}
Owing to the causality and the advantage on training as explained in the previous subsection, uni-directional LSTM are often used in real-time speech enhancement \cite{Naithani2017lowLatency,Tan2018convRNN,Pandey2019TCNN}.
While LSTM performs well in practice, any causal RNN written in the form of Eq.~\eqref{eq:causalTFmask} can be used for real-time speech enhancement.
That is, it should be possible to construct a computationally cheaper RNN for the real-time application if the training issue can be solved.

\begin{figure}[t]
  \centering
  \includegraphics[width=0.95\columnwidth]{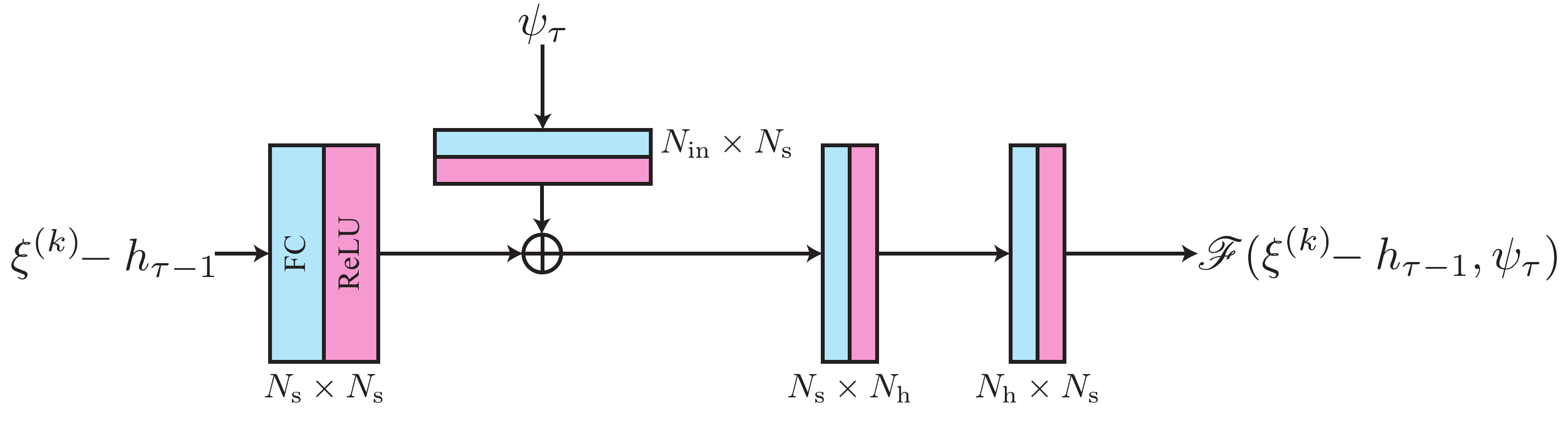}
  \vspace{-8pt}
  \caption{Illustration of DNN $\mathscr{F}$ utilized for ERNN in this paper. ``FC" stands for a fully-connected layer, and $N_{\rm s}$ and $N_{\rm h}$ are the dimension of the matrix of each fully-connected layer.}
  \label{fig:miniRNNArch}
\end{figure}

\mysection{Proposed method}
For real-time speech enhancement, a causal DNN with fewer parameters is preferred for reducing the computational requirement.
Considering such conditions, we propose a causal DNN-based speech enhancement method using ERNN illustrated in Fig.~\ref{fig:LSTM_ERNN}(b).

\mysubsection{Equilibriated recurrent neural network (ERNN)}
ERNN is an RNN which avoids the vanishing/exploding gradient problem by the skip connections and repeated application of the same block \cite{kag2019rnns}.
It is inspired by the fixed point recursion of the implicit discretization scheme for an ordinary differential equation.
By introducing an intermediate variable $\xi^{(k)}$ with iteration index $k=0,\ldots,K-1$, a simple form of ERNN can be written as
\begin{equation}
    \xi^{(k+1)} = \xi^{(k)} + \eta^{(k)}[\mathscr{F}(\psi_\tau,\xi^{(k)}+h_{\tau-1}) - (\xi^{(k)}+h_{\tau-1})],
\label{eq:ernn_iter}
\end{equation}
where $\eta^{(k)}$ is a small trainable scalar, $K$ is the total number of iteration, the initial value $\xi^{(0)}$ is typically $0$, and the updated state vector $h_\tau$ is given as the iterated result $h_\tau = \xi^{(K)}$, i.e., ERNN returns $h_\tau$ after $K$ iteration based on the inputs $\psi_\tau$ and $h_{\tau-1}$ as in Eq.~\eqref{eq:simplestRNN}.
Here, $\mathscr{F}$ is a nonlinear function implemented by a neural network, which makes Eq.~\eqref{eq:ernn_iter} a multilayer RNN as in Fig.~\ref{fig:LSTM_ERNN}(b).

The notable property of ERNN is that its gradient does not vanish or explode in the ideal situation \cite{kag2019rnns}.
That is, the norm of the gradient is equal to one: $\|\partial h_c/\partial h_p\| = 1\;\;(p<c)$.
Therefore, it is expected that ERNN can learn the long-term dependencies without suffering from the training issue because the gradient survives in the parameter update for all time instances.
This property should allow us to simplify the network because the gated units used in LSTM are not necessary anymore for alleviating the difficulty of training.
We experimentally show later in the next section that a simple ERNN with much fewer parameters can compete with LSTM.

\mysubsection{Proposed speech enhancement method using ERNN}
We propose a speech enhancement method based on a causal ERNN.
The proposed method estimates the T-F mask for the current time frame $G_\tau$ by ERNN whose input is based only on the current input feature $\psi_\tau$ and the hidden state vector $h_{\tau-1}$ as
\begin{equation}
    G_\tau = \sigma(Wh_{\tau}+b),\qquad
    h_\tau = \mathrm{ERNN}_K^\mathscr{F}\!(\psi_{\tau},h_{\tau-1}),
\end{equation}
where $W$ and $b$ are the matrix and bias of the fully-connected layer, respectively, $\sigma(\cdot)$ is the sigmoid function, and $\mathrm{ERNN}_K^\mathscr{F}\!(\cdot,\cdot)$ is the function iterating Eq.~\eqref{eq:ernn_iter} $K$ times from an initial value $\xi^{(0)}$ using the nonlinear function $\mathscr{F}$.

Since the expressive power of ERNN is determined by the nonlinear function $\mathscr{F}$, the performance and the degree of computational requirements can be traded by appropriately designing it.
For real-time applications, we aim to reduce the number of parameters of the system while maintaining the performance so that the proposed method can compete with the standard LSTM-based methods.
As the first step of the investigation of the proposed method, we consider a fully-connected DNN with the ReLU activation as illustrated in Fig.~\ref{fig:miniRNNArch} because it is easy to adjust the number of parameters by changing the size of the fully-connected layers.
As the DNN $\mathscr{F}$ is common for all $k$ in the iteration of Eq.~\eqref{eq:ernn_iter}, the number of parameters is that of $\mathscr{F}$ plus $K$ (comes from the scalers $\eta^{(0)},\ldots,\eta^{(K-1)}$).
Note that this choice of $\mathscr{F}$ is merely an example, and it should be possible to design a better network consisting of fewer parameters.

\mysection{Experiment}
In order to confirm the effectiveness of the proposed method, the performance of speech enhancement was investigated by comparing with LSTM-based methods as the baselines.
We conducted two experiments.
As the first experiment, we compared the performance and the number of parameters of the proposed and conventional methods by selecting the same number of the cell units.
In the second experiment, the number of parameters of the proposed method was decreased to see how the performance varies for smaller DNN.
Our implementation of these experiments is openly available online\footnote{
\url{https://github.com/dtake1336/ERNN-for-speech-enhancement}
}.

\mysubsection{Experimental condition}
\vspace{7pt}

\begin{table}[t]
\centering
\caption{Network architectures for the experiment.}
\vspace{2pt}
\begin{tabular}{l | c | l} \hline \hline
Layer & Type & Size (activation) \\ \hline
\multicolumn{3}{c}{LSTM2/BLSTM2}\\ \hline
Layer1 & LSTM/BLSTM & 257         $\to$ $N_{\rm s}$ \\ 
Layer2 & LSTM/BLSTM & $N_{\rm s}$ $\to$ $N_{\rm s}$ \\ 
output & Fully      & $N_{\rm s}$ $\to$ 257 (sigmoid) \\ \hline
\multicolumn{3}{c}{ERNN}\\ \hline
Layer1 & ERNN  & 257         $\to$ $N_{\rm s}$ \\ 
output & Fully & $N_{\rm s}$ $\to$ 257 (sigmoid) \\ 
\hline \hline
\end{tabular}
\label{tab:DNNachi}
\vspace{-10pt}
\end{table}

\mysubsubsection{Dataset}
We utilized the VoiceBank-DEMAND dataset constructed by Valentini 
{\it et al}.~\cite{valentini2016investigating} which is openly available%
\footnote{\url{http://dx.doi.org/10.7488/ds/1356}} and frequently used in the literature of DNN-based speech enhancement.
It consists of train set and test set which contain noisy mixtures and clean speech signals, respectively, i.e., noise and speech signals were already mixed by the authors \cite{valentini2016investigating}.
They consist of 28 and 2 speakers (11\,572 and 824 utterances) \cite{veaux2013voice} which are contaminated by 10 (DEMAND, speech-shaped noise, and babble) and 5 types of noise (DEMAND) \cite{thiemann2013diverse}, respectively.
All data were downsampled from 48 kHz to 16 kHz.

\mysubsubsection{DNN architecture, loss function and training setup}
The parameters of STFT were the 512 points (32 ms) Hann window, 256 points time-shifting, and 512 points FFT length, and the inverse STFT was implemented by its canonical dual \cite{yatabe2019DGT}. 
In the proposed method, DNN $\mathscr{F}$ in ERNN was that illustrated in Fig.~\ref{fig:miniRNNArch}, and the iteration number $K$ was varied as 1/3/5. 
The size of hidden vector $N_{\rm s}$ and $N_{\rm h}$ were varied as 512/256 and 512/256/128/64/32, respectively.
For the baseline methods, two-layered LSTM and BLSTM, which are popular and have been successfully applied to speech enhancement \cite{Erdogan2015phase}, consisting of 512/256 cells were used as summarized in Table~\ref{tab:DNNachi}.
For the input feature, log-magnitude spectrogram,
\begin{equation}
\Psi_{\omega, \tau} = {\rm ln}(|X_{\omega, \tau}|),
\end{equation}
was used for all networks, where $|\cdot|$ denotes the absolute value.
As an activation function of the output layer, the sigmoid function was used for limiting the values within the range of 0 to 1.

For the loss function in the training, the mean absolute error measured in the time domain was used:
\vspace{-4pt}
\begin{equation}
     \mathcal{J}_{\rm MAE}(\theta) = 
     \frac{1}{T}
     \sum_{t = 1}^{T}
     \,\bigl|\, s_t - 
     {\rm iSTFT}(\mathcal{M}_{\theta}(\Psi)\odot X)_t
     \,\bigr|,
     \label{eq:maeLoss}
     \vspace{-4pt}
\end{equation}
where $\odot$ is the element-wise multiplication, and ${\rm iSTFT}(\cdot)$ denotes the inverse STFT.
Each DNN was trained 200 epochs where each epoch contained 11\,572 utterances.
A one-second-long segment was randomly picked up for each utterance, and mini-batch size was 16.
Adam optimizer \cite{Kingma2015} was utilized with a fixed learning rate 0.0001.

The performance of speech enhancement was measured by PESQ\cite{wPESQ} and three  measures CSIG, CBAK, and COVL~\cite{Hu2008eval} which are the popular predictor of the mean opinion score~(MOS) of the signal distortion, the background noise interference, and the overall effect, respectively.

\begin{table}[t]
\centering
\caption{Results for comparison $(N_{\rm s} = 256)$}
\vspace{2pt}
\begin{tabular}{ 
@{}C{0.155}@{} | @{}C{0.08}@{}  | @{}C{0.08}@{}  | @{}C{0.045}@{}| @{}C{0.14}@{}  |
@{}C{0.125}@{} | @{}C{0.125}@{} | @{}C{0.125}@{} | @{}C{0.125}@{} } 
\hline \hline
DNN            & $N_{\rm s}$    & $N_{\rm h}$    & $K$     & Params. &
PESQ           & CSIG           & CBAK           & COVL \\ 
\hline
       &     &     & 1 &      & 2.42 & 3.57 & 2.58 & 2.98 \\
\cline{4-4} \cline{6-9} 
ERNN   &     & 256 & 3 & 329k & 2.49 & 3.58 & $\bm{2.62}$ & 3.02 \\
\cline{4-4} \cline{6-9} 
       & 256 &     & 5 &      & 2.43 & 3.56 & 2.58 & 2.98 \\
\cline{1-1} \cline{3-9} 
BLSTM2 &     & \multirow{2}*{--} & \multirow{2}*{--} &  2.76M & $\bm{2.50}$ & $\bm{3.65}$ & $\bm{2.62}$ & $\bm{3.06}$ \\
\cline{1-1} \cline{5-9} 
LSTM2  &     &    &  & 1.12M & 2.34 & 3.49 & 2.55 & 2.90 \\\hline
\hline 
\end{tabular}
\label{tab:res1dim256}
\vspace{2pt}
\caption{Results for comparison $(N_{\rm s} = 512)$}
\vspace{2pt}
\begin{tabular}{ 
@{}C{0.155}@{} | @{}C{0.08}@{}  | @{}C{0.08}@{}  | @{}C{0.045}@{}| @{}C{0.14}@{}  |
@{}C{0.125}@{} | @{}C{0.125}@{} | @{}C{0.125}@{} | @{}C{0.125}@{} } 
\hline \hline
DNN            & $N_{\rm s}$    & $N_{\rm h}$    & $K$     & Params. &
PESQ           & CSIG           & CBAK           & COVL \\ 
\hline
       &     &     & 1  &       & 2.43 & 3.65 & 2.60 & 3.03 \\ 
\cline{4-4} \cline{6-9} 
ERNN   &     & 512 & 3  & 1.05M & 2.43 & 3.60 & 2.59 & 3.00 \\ 
\cline{4-4} \cline{6-9} 
       & 512 &     & 5  &       & 2.41 & $\bm{3.67}$ & 2.58 & 3.02 \\ 
\cline{1-1} \cline{3-9} 
BLSTM2 &     & \multirow{2}*{--}  & \multirow{2}*{--} & 9.72M & $\bm{2.53}$ & $\bm{3.67}$ & $\bm{2.65}$ & $\bm{3.08}$ \\
\cline{1-1} \cline{5-9} 
LSTM2  &     &   &  & 3.81M & 2.45 & 3.63 & 2.61 & 3.03 \\ \hline
\hline 
\end{tabular}
\label{tab:res1dim512}
\vspace{-8pt}
\end{table}

\begin{table}[t]
\centering
\caption{Results of varying $N_{\rm h}$ $(N_{\rm s}= 256)$}
\vspace{3pt}
\begin{tabular}{ 
@{}C{0.155}@{} | @{}C{0.08}@{}  | @{}C{0.08}@{}  | @{}C{0.045}@{}| @{}C{0.14}@{}  |
@{}C{0.125}@{} | @{}C{0.125}@{} | @{}C{0.125}@{} | @{}C{0.125}@{} } 
\hline \hline
DNN & $N_{\rm s}$ & $N_{\rm h}$ & $K$ & Params. & PESQ & CSIG & CBAK & COVL \\ \hline
\multirow{9}*{ERNN} & \multirow{9}*{256} 
 &     & 1 &     &2.30&3.34&2.52&2.80 \\
\cline{4-4}\cline{6-9}
&& 32  & 3 & 215k&2.39&3.54&2.57&2.95 \\
\cline{4-4}\cline{6-9}
&&     & 5 &     &2.32&3.45&3.45&2.87 \\
\cline{3-9}
&&     & 1 &     &2.43&3.60&2.59&3.00 \\
\cline{4-4}\cline{6-9}
&& 64  & 3 & 231k&2.40&3.56&2.58&2.97 \\
\cline{4-4}\cline{6-9}
&&     & 5 &     &2.37&3.57&2.56&2.96 \\
\cline{3-9}
&&     & 1 &     &2.45&3.64&2.61&3.03 \\
\cline{4-4}\cline{6-9}
&& 128 & 3 & 264k&2.40&3.57&2.58&2.97 \\
\cline{4-4}\cline{6-9}
&&     & 5 &     & $\bm{2.49}$&$\bm{3.71}$& $\bm{2.63}$& $\bm{3.09}$ \\
\hline \hline 
\end{tabular}
\label{tab:resultNh256}

\vspace{3pt}

\caption{Results of varying $N_{\rm h}$ $(N_{\rm s}= 512)$}
\vspace{3pt}
\begin{tabular}{ 
@{}C{0.155}@{} | @{}C{0.08}@{}  | @{}C{0.08}@{}  | @{}C{0.045}@{}| @{}C{0.14}@{}  |
@{}C{0.125}@{} | @{}C{0.125}@{} | @{}C{0.125}@{} | @{}C{0.125}@{} } 
\hline \hline
DNN & $N_{\rm s}$ & $N_{\rm h}$ & $K$ & Params. & PESQ & CSIG & CBAK & COVL \\ \hline
\multirow{12}*{ERNN} & \multirow{12}*{512} 
 &     & 1 &     &2.35&3.43&2.54&2.87 \\
\cline{4-4}\cline{6-9}
&& 32  & 3 & 560k&2.40&3.58&2.58&2.98 \\
\cline{4-4}\cline{6-9}
&&     & 5 &     &2.41&3.62&2.58&3.00 \\
\cline{3-9}
&&     & 1 &     &2.44&3.56&2.59&2.98 \\
\cline{4-4}\cline{6-9}
&& 64  & 3 & 593k&2.47&3.60&2.61&3.02 \\
\cline{4-4}\cline{6-9}
&&     & 5 &     &2.45&3.63&2.61&3.03 \\
\cline{3-9}
&&     & 1 &     &2.49&3.70&2.62&3.08 \\
\cline{4-4}\cline{6-9}
&& 128 & 3 & 658k&2.36&3.52&2.56&2.93 \\
\cline{4-4}\cline{6-9}
&&     & 5 &     &2.52&3.69&2.64&3.09 \\
\cline{3-9}
&&     & 1 &     &2.52&3.68&2.63&3.09 \\
\cline{4-4}\cline{6-9}
&& 256 & 3 & 790k&2.48&$\bm{3.75}$&2.63&3.10 \\
\cline{4-4}\cline{6-9}
&&     & 5 &     & $\bm{2.54}$&3.74& $\bm{2.65}$&$\bm{3.13}$ \\
\hline \hline
\end{tabular}
\label{tab:resultNh512}
\vspace{-0pt}
\end{table}

\mysubsection{Results}

The results for comparison are summarized in Tables~\ref{tab:res1dim256}~and~\ref{tab:res1dim512}, where the cell sizes were 256 and 512, respectively.
As well known in the speech enhancement literature, BLSTM performed better than LSTM because BLSTM is non-causal and can use the information from the future, while LSTM is causal and can only use the past information.
Since BLSTM cannot be utilized for real-time applications, its scores are merely a reference, and LSTM is the direct competitor of the proposed method.
Comparing with LSTM, the proposed method obtained almost the same performance in every situation.
For some situations, the proposed method also obtained the similar performance compared to BLSTM even though the proposed method is causal and contains about 1/9 parameters.
This should be because ERNN was able to successfully learn the long-term dependencies of the speech signals.

Since our aim is to construct a network with fewer parameters, the number of the parameters of the proposed method was reduced by changing the dimension of the linear layer $N_{\rm h}$ (see Fig.~\ref{fig:miniRNNArch}).
The results are summarized in Tables~\ref{tab:resultNh256}~and~\ref{tab:resultNh512}, where the result for $N_{\rm s}=N_{\rm h}$ can be found in Tables~\ref{tab:res1dim256}~and~\ref{tab:res1dim512}.
As a general tendency, reducing the number of parameters gradually degrades the performance.
However, the amount of degradation is not so significant, which indicates that the proposed method can reduce the computational requirement without losing the performance much.
In terms of the number of iteration $K$, more iteration tends to slightly improve the performance.
The proposed method can reduce the computational requirement by reducing $K$, where $K=1$ means that the network $\mathscr{F}$ is applied only once at each time frame.

Note that, by comparing the best scores in Table~\ref{tab:resultNh256} with LSTM in Table~\ref{tab:res1dim512}, the proposed method outperformed LSTM with less than 1/14 parameters.
It is also comparable to BLSTM in Table~\ref{tab:res1dim512} with less than 1/36 parameters and BLSTM in Table~\ref{tab:res1dim256} with around 1/10 parameters.
Again, BLSTM cannot perform in real time as it is non-causal, and thus the proposed method should be compared with LSTM.
While LSTM lost noticeable amount of performance by reducing the parameters as shown in Tables~\ref{tab:res1dim256}~and~\ref{tab:res1dim512}, the proposed method can reduce the number of parameters with moderate amount of degradation of the performance as in the tables.
Therefore, we confirmed the effectiveness of the proposed method in real-time speech enhancement as it can be performed by much lower computational cost compared to the standard LSTM networks.

\mysection{Conclusions}
\vspace{1pt}
In this paper, the causal DNN-based speech enhancement method using ERNN was proposed for real-time applications.
By using ERNN, the number of parameters can be decreased thanks to its ability of learning the long-term dependencies without the vanishing gradient problem.
The experimental results indicated that, while the standard LSTM lost the performance by reducing the number of parameters, the proposed method can effectively trade the performance and computational requirement which is preferable for performing speech enhancement on resource-limited devices.
As this paper only considered a simple fully-connected DNN with ReLU activation as an example for ERNN, our future works include investigation of a better network performing well with less number of parameters.

\end{document}